\documentclass[sigconf]{acmart} 

\usepackage{soul}

\AtBeginDocument{%
  \providecommand\BibTeX{{%
    \normalfont B\kern-0.5em{\scshape i\kern-0.25em b}\kern-0.8em\TeX}}}

\copyrightyear{2021}
\acmYear{2021}
\setcopyright{acmlicensed}\acmConference[CHI '21]{CHI Conference on Human Factors in Computing Systems}{May 8--13, 2021}{Yokohama, Japan}
\acmBooktitle{CHI Conference on Human Factors in Computing Systems (CHI '21), May 8--13, 2021, Yokohama, Japan}
\acmPrice{15.00}
\acmDOI{10.1145/3411764.3445385}
\acmISBN{978-1-4503-8096-6/21/05}

\begin{document}

\title[Designing AI for Time-Constrained Medical Decisions]{Designing AI for Trust and Collaboration in Time-Constrained Medical Decisions: A Sociotechnical Lens}


\author{Maia Jacobs}
\orcid{0000-0001-8500-0277}
\affiliation{Northwestern University}
\email{maia.jacobs@northwestern.edu}

\author{Jeffrey He}
\affiliation{Harvard University}
\email{jdhe@college.harvard.edu}

\author{Melanie F. Pradier}
\affiliation{Microsoft Research}
\email{melanief@microsoft.com}

\author{Barbara Lam}
\affiliation{Beth Israel Deaconess Medical Center}
\email{blam@bidmc.harvard.edu}

\author{Andrew C. Ahn}
\affiliation{Harvard Medical School}
\affiliation{Beth Israel Deaconess Medical Center}
\email{aahn@bidmc.harvard.edu}

\author{Thomas H. McCoy}
\affiliation{Massachusetts General Hospital}
\affiliation{Harvard Medical School}
\email{tmccoy@mgh.harvard.edu}

\author{Roy H. Perlis}
\affiliation{Massachusetts General Hospital}
\affiliation{Harvard Medical School}
\email{rperlis@mgh.harvard.edu}

\author{Finale Doshi-Velez}
\affiliation{Harvard University}
\email{finale@seas.harvard.edu}

\author{Krzysztof Z. Gajos}
\affiliation{Harvard University}
\email{kgajos@g.harvard.edu}

\renewcommand{\shortauthors}{Jacobs, et al.}

\begin{abstract}
Major depressive disorder is a debilitating disease affecting 264 million people worldwide. While many antidepressant medications are available, few clinical guidelines support choosing among them. Decision support tools (DSTs) embodying machine learning models may help improve the treatment selection process, but often fail in clinical practice due to poor system integration.
  
We use an iterative, co-design process to investigate clinicians' perceptions of using DSTs in antidepressant treatment decisions. We identify ways in which DSTs need to engage with the healthcare sociotechnical system, including clinical processes, patient preferences, resource constraints, and domain knowledge. Our results suggest that clinical DSTs should be designed as multi-user systems that support patient-provider collaboration and offer on-demand explanations that address discrepancies between predictions and current standards of care. Through this work, we demonstrate how current trends in explainable AI may be inappropriate for clinical environments and consider paths towards designing these tools for real-world medical systems.


  

\end{abstract}




\maketitle

\section{Introduction}
Advances in artificial intelligence (AI) and machine learning (ML) offer opportunities to uncover complex data patterns. In medicine, the integration of AI tools could lead to a substantial paradigm shift in which human-AI collaboration becomes integrated in medical decision-making. Such AI-powered decision support tools (DSTs) may support many clinical practices, including diagnosing illnesses, selecting the optimal treatment for a patient, and predicting disease trajectories \cite{Yang2015}. While the promise of AI in medicine is alluring, the use of these systems in healthcare has been discussed for decades, and yet few of these tools have resulted in successful implementation and use in clinical practice~\cite{Middleton2016}.

One area of healthcare that researchers have expected to benefit from the implementation of DSTs, but has yet to adopt such technological support, is major depressive disorder (MDD). Antidepressant medications are a common form of treatment for MDD, but selecting an effective treatment for a patient is a complex task. The majority of mental health care is initiated in primary care settings, yet the extent of training primary care providers (PCP) receive in managing MDD can vary widely \cite{Thielke2007, Wolf2008}, and contemporary treatment guidelines provide little support in choosing among them \cite{Kennedy2016}. Limited guidelines as well as heterogeneity in patients' symptoms and in patients' tolerability of antidepressants often results in using trial and error to identify an effective treatment \cite{Trivedi2008}. Currently, an estimated one-third of patients fail to reach remission even after four antidepressant trials \cite{Rush2006}. The frequency of ineffective drug trials has resulted in the psychiatry community calling for more information on which treatments will be most effective for an individual patient \cite{Simon2010}. In response, we have seen several studies focused on creating ML models to support MDD treatment decisions \cite{Shatte2019, Pradier2020, Hughes2018}.



While ML models for antidepressant treatment selection exist, these systems are rarely integrated into clinical practice due to low user acceptance and a failure to account for user expectations in the system design \cite{Khairat2018a, Fusar-Poli2018}. Therefore, motivated by Berg's sociotechnical approach, which highlights the importance of empirical research of healthcare practices in which the technology will be used \cite{Berg1999}, we consider the social, technical, and organizational issues that must be considered in the design of DSTs for MDD treatment selection.


Using an iterative design process and two qualitative studies with PCPs, our findings raise many challenges to integrating ML-enabled tools into real clinical workflows. We found that intelligent decision support tools will need to seamlessly integrate into clinicians' time constrained workflows, which typically involve short appointments with patients to understand their symptoms and make treatment decisions. We also found that clinicians' busy schedules influenced how they thought about trust as it relates to black-box ML models. Conversations revealed that PCPs wanted DSTs that: 1)~engage patients in decision-making, 2)~connect DST output to existing healthcare system processes, 3)~do not require making decisions of trust at every interaction, and 4)~compare and contrast DST output with existing standards of care. 

Based on these findings, we discuss how using a sociotechnical lens challenges current trends in the design of explainable AI. We highlight the need to design DSTs as multi-user systems that facilitate patient-provider-AI collaboration. Further, our work reveals issues with using explanations that encourage users to make determinations of trust for each prediction. We recommend that for time-constrained medical environments, we shift from designing explanations for every decision to on-demand  explanations that contrast AI recommendations with current standards of care. 

While we have seen many recent advances related to the use of HCI methods for designing AI tools, we have seen few studies that discuss what it will take to make these tools work within complex sociotechnical systems. Our contributions include the following:




\begin{enumerate}
    \item We use an iterative design process to create a prototype of an MDD decision support tool that integrates both patient-level prognostic predictions and treatment selection support.
    \item Based on primary care providers' feedback, we present the important facets of the healthcare sociotechnical system that must be considered in the design and development of machine learning tools for real-world clinical care.
    \item We discuss how using a sociotechnical lens presents new opportunities and challenges for both HCI and ML research. 
\end{enumerate}

\section{Related Work}
\subsection{Co-designing AI Systems}
In recent years, we have seen increased interest in embedding AI tools in a variety of contexts, such as the justice system \cite{Albright2019, Stevenson2018}, the U.S. child welfare system \cite{Saxena2020}, and medicine \cite{Wiens2019,Sitapi2016,He2019}. Past studies have identified a number of problems with implementing existing models into real-world workflows. One issue is that recent AI work has focused on improving the accuracy of the model, rather than the needs of the intended user \cite{Selbst2019}, and improving model accuracy does not always correlate with overall performance once implemented in the real-world \cite{Bansal2019, Beede2020}. In the context of antidepressant treatment selection, Jacobs et al. found in a factorial experiment that AI recommendations did not improve treatment selection accuracy, highlighting the need for research that engages directly with clinicians to create tools that are interpretable and useful \cite{Jacobs2021}.

In the past few years, we have seen an upswing in CHI research that uses co-design methods to consider the real-world challenges, beyond accuracy, that must be considered in the design and development of these tools. HCI research has helped to advance the field of explainable AI, examining how people interact with AI tools and designing tools to help end-users understand the inner workings of machine learning models \cite{Yin2019a, Bucinca2020, Cai2019, Hohman2019, Amershi2019, Oh2018}. We have also seen an increased use of co-design activities that give end-users a greater voice in how these systems should function \cite{Luo2019, Bussone2015, Tonekaboni2019, Yang2019a}. This research has revealed the importance of context in determining if and when end-users want computational assistance.

We assert that another problem is the lack of context awareness of the broader sociotechnical systems in which these tools are being embedded. In a recent paper, Selbst et al. discuss how a failure to use a sociotechnical lens can lead ML tools to be ineffective \cite{Selbst2019}. Beede et al. also recently showed how environmental factors in a clinical setting influenced the usability of a deep learning system \cite{Beede2020}. Sociotechnical approaches have been important in healthcare for understanding how new technologies may be effectively integrated into the social processes that make up healthcare work \cite{Berg2003}. Here we use clinical perspectives to consider the broader sociotechnical context that will be necessary to consider when creating AI systems for real-world use.

\subsection{Clinical Decision Support Tools: Healthcare System Integration}
Decision support tools are computational systems created to facilitate medical decision-making \cite{Osheroff2007}. These tools can be designed to provide a range of outputs, including treatment recommendations, prognosis predictions, and patient diagnoses \cite{Yang2015}. Clinical DSTs have long attracted researchers due to their ability to perform tasks that exceed human capabilities, such as extracting information and patterns from large amounts of data. These data-driven tools offer the opportunity to improve health outcomes and reduce human errors in the decision-making process \cite{Middleton2016}.

Despite many years of enthusiasm towards these technologies~\cite{Osheroff2007, Middleton2016}, the vast majority of these tools fail once they are deployed in real-world health systems. 
Notably, DSTs are typically not abandoned due to poor performance, but rather due to failures in accounting for the complexity of the healthcare sociotechnical system~\cite{Pratt2004c}. Researchers have found that poor workflow integration, low context awareness, and a failure to incorporate clinicians' expectations have led to low user acceptance of DSTs \cite{Middleton2016, Sittig2008, Pratt2004c, Khairat2018a}. Several papers have discussed the relationship between poor DST integration and low user acceptance. For example, studies in medical informatics found that when DST alerts arise at inopportune times (from clinicians' vantage), the alerts become ignored or overridden \cite{Nanji2014, Bryant2014, Kuperman2007}. In response, there have been calls both within and outside of the field of HCI to use CSCW and HCI methods to improve these tools \cite{Pratt2004c, Sittig2008, Yang2015}.

Within HCI, recent work has helped to establish clinical expectations for DSTs and develop guidelines for improving clinician-AI interactions. For example, Cai et al. found that refinement tools were considered more helpful and easier to use than traditional tools when interacting with image retrieval systems \cite{Cai2019}. In another study looking at image retrieval tools, Xie et al. developed a set of design recommendations for supporting clinician exploration and subsequent understanding of AI tools \cite{Xie2020}. Finally, Yang et al. designed a DST prototype to facilitate artificial heart implant decisions \cite{Yang2019a}. Using concepts from unremarkable computing, they aimed to make AI prediction unobtrusive in clinicians' workflow and found that this integration supported clinicians' acceptance of the technology. These studies  collectively provide insights into how DSTs may be better situated into clinical routines. Through conversations with clinicians we have found that successful DST implementation will also require a broader context awareness. We extend existing literature by identifying other aspects of the healthcare system that must be considered in the creation and deployment of novel DSTs.

\subsection{Major Depressive Disorder}
Major depressive disorder (MDD) is a brain disorder characterized by depressed mood, loss of interest in daily activities, as well as change in associated symptoms such as sleep, energy, eating, concentration, and thoughts of death or suicide. Lifetime prevalence of MDD in the United States is estimated to exceed 15\% \cite{Kessler2003}. Treatments for MDD supported by randomized, controlled trials include antidepressant medications, cognitive-behavioral therapy, and somatic therapies such as transcranial magnetic stimulation and electroconvulsive therapy \cite{Fava200}.

Finding an effective antidepressant medication for someone diagnosed with MDD is an important but difficult task. 
Both mental health specialists and PCPs (including physicians and nurse practitioners) are authorized to write prescriptions. The majority of mental healthcare is provided within primary care settings \cite{Wolf2008}. However, primary care appointments last an average of only 20 minutes in the United States, and internationally primary care appointments can be as short as a few minutes \cite{Irving2017}. Such short encounters are considered insufficient for effective mental health care \cite{Hutton2007}. Further, the training that PCPs receive on antidepressant treatment selection can be highly variable \cite{Thielke2007, Wolf2008}. 

To support treatment selection, the Canadian Network for Mood and Anxiety Treatments provides widely followed treatment guidelines for 25 antidepressants, organized as first-, second-, and (a small number of) third-line treatments \cite{Kennedy2016}. First-line treatments are the recommended initial treatment options. If ineffective, the provider may try second- and then third-line treatment, which often have more severe side effects or drug interactions. While these treatment guidelines provide a useful resource, the heterogeneity in patients' symptoms and tolerability of antidepressants means that identifying an effective treatment for an individual remains a process of trial and error~\cite{Trivedi2008}. 

The trial and error involved in identifying effective treatment for a given individual has prompted calls for more integration of evidence-based medicine in the treatment of mental health disorders~\cite{Simon2010, Menke2018, Perlis2016}. Current state of the art models provide prognostic predictions and support treatment selection \cite{Pradier2020, Hughes2018}. A number of questions must be answered before implementing these models, particularly regarding clinician expectations and system integration. To support the future integration of DSTs into primary care, we seek to better understand PCP's support needs and expectations for how DSTs will function in the healthcare system.

\section{USER STUDY 1: NEEDS ELICITATION}

\subsection{Methods}
\subsubsection{Participants and Recruitment}
We worked with primary care physicians who currently prescribe antidepressant treatments. We recruited participants at a large academic medical center. We sent information about the study to an email list of primary care providers (including physicians, nurse practitioners, and residents). We provided each participant with a \$20 Amazon gift card to thank them for their time. 

\subsubsection{Study Design}
This study was approved by the Harvard Institutional Review Board. We used semi-structured interviews and focus groups with no more than two participants. Each session lasted 30 minutes due to participants' busy schedules. We used the first 15 minutes to discuss existing decision-making processes and the second 15 minutes to discuss future state ideas. To drive discussion on current-state processes we used four questions to guide the semi-structured interview: 1) What factors do you consider when selecting an antidepressant? 2) What do you do if you are unsure which antidepressant to select? 3) Where is there room for improvement in this process? 4) Is there any information you wish you had when selecting a treatment for a patient?

We also wished to include clinicians in co-design activities to discuss their expectations for future DSTs. As this study was scheduled for March 2020, all study activities were revised and done remotely using Zoom. Prior work has discussed the challenges of running remote design studies \cite{MacLeod2017}. To facilitate future-state ideation, we designed low fidelity prototypes to demonstrate a variety of features and information we could potentially derive based on existing ML research that uses electronic medical record data. The prototypes included the following features. 

\begin{figure*}[]
  \centering
  \includegraphics[width=.86\textwidth]{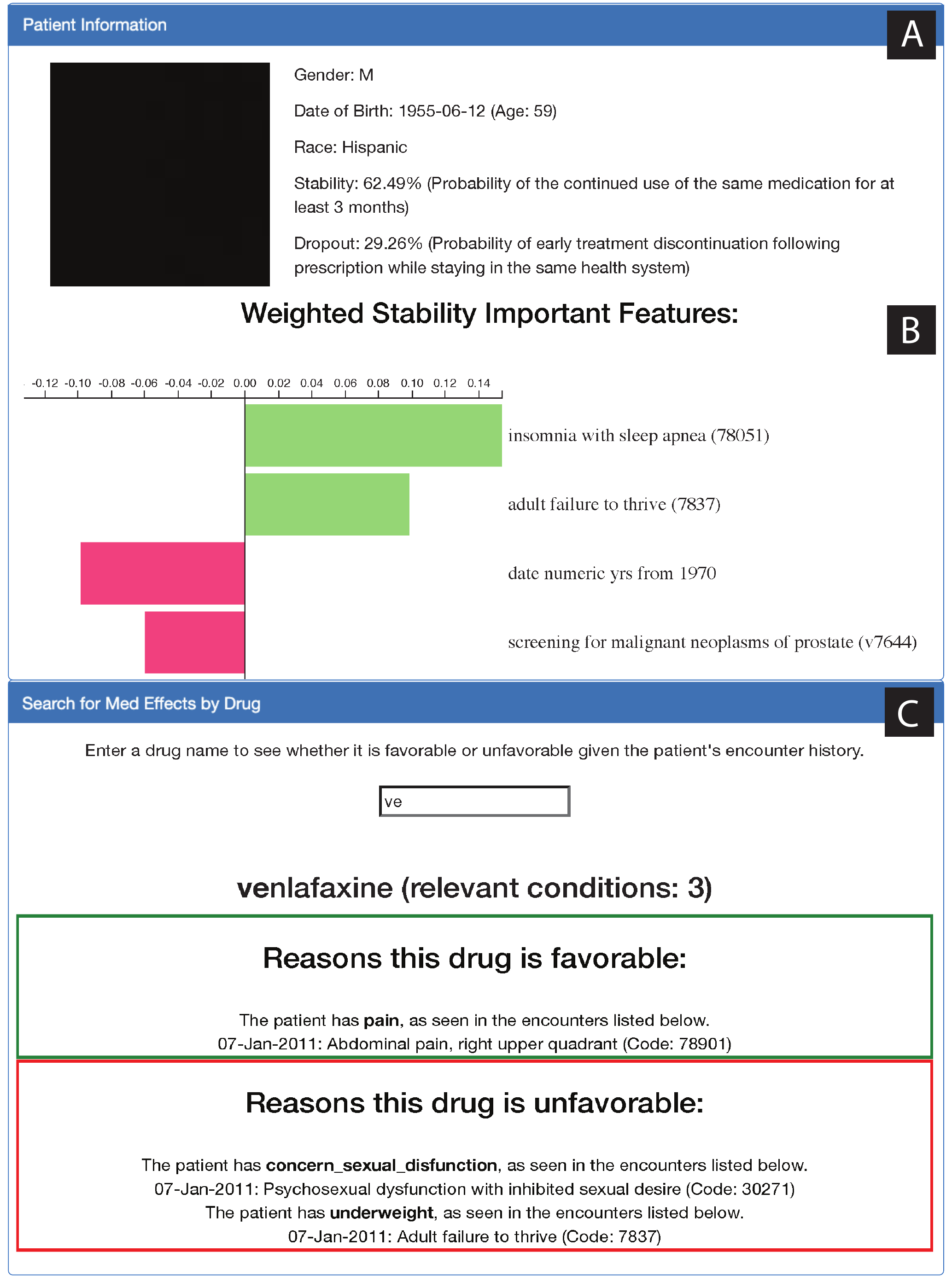}
  \caption{Features included in the initial prototype (from top to bottom): (A) a patient scenario with stability and dropout scores, (B) stability score feature importance explanation, (C) personalized treatment recommendations.}
  \Description{The figure shows three features included in the prototype. The top image includes a list of patient information, a stability score of 62.49\% and dropout score of 29.26\%. The middle image shows a feature importance explanation titled 'Weighted stability Important Features'. The bottom image shows an treatment selection screen in which providers may search for a treatment and see associated rules that are relevant to the patient, explaining why the treatment may or may not be favorable.}
  \label{fig:one}
\end{figure*}

\textbf{Patient-level prognostic predictions:} 
\begin{enumerate}
    \item Stability score (figure \ref{fig:one}A): The probability of continued use of the same medication for at least 3 months \cite{Hughes2020}.
    \item Dropout score (figure \ref{fig:one}A): The probability of early treatment discontinuation following prescription while staying in the same health system \cite{Pradier2020}. 
    \item Stability and dropout feature importance (figure \ref{fig:one}B): A widely used approach for explaining ML predictions that shows relevant features and their contributions to the prediction \cite{Guidotti2018}. In this context, these features include electronic health record codes that contributed to the stability and dropout predictions. 
\end{enumerate}

\textbf{Treatment selection support:}
\begin{enumerate}
    \item Personalized treatment recommendations (figure \ref{fig:one}C): To support treatment selection, we used drug-specific rules based on drug interactions. The rules were curated by two collaborating psychopharmacologists with a mean of 12 years of clinical practice in the United States. Example rules include \textit{for anxiety favor mirtazapine} and \textit{for poor concentration favor bupropion}.   
    The interface shows personalized treatment recommendations by showing which rules are relevant to a patient based on their electronic medical records. Through these rules, the tool can show which treatments would be favorable or not favorable for a patient, based on their medical history and the drug side effects.
\end{enumerate}

The prototype includes a fabricated patient scenario, as seen in figure \ref{fig:one}A. The scenario was vetted by psychopharmacology experts as a reasonable example of a real-world patient. In the user study, we asked participants to reflect on these prototypes by considering which aspects they found helpful or unhelpful, what they would change, and other ideas for how technology may support antidepressant treatment selection.

\subsubsection{Data Analysis}
All interviews were audio recorded and transcribed. To analyze the data we used a grounded theory approach~\cite{Charmaz2012}. The first author open coded the transcripts, comparing and contrasting existing decision-making processes and prototype feedback, and looking for patterns in the dataset. The research team then met to discuss emerging themes and review associated data segments. Using the established themes, the first author re-coded the transcripts and met with team members to discuss new categories as they emerged in subsequent coding iterations and to validate the final themes.

\subsection{Findings: Clinical Expectations for ML Decision Support}
Ten PCPs volunteered to participate in this study (six physicians, three resident physicians, and one nurse practitioner). Participants had an average of 12.5 years of experience prescribing antidepressants (SD=12.1). Table \ref{tab:participants} includes participant details across both of the user studies described in this paper, including participants' years of experience prescribing antidepressants, and in which of the two studies they participated. 

Through the interviews participants affirmed many established treatment selection challenges. Most commonly, participants spoke about their limited familiarity with antidepressant treatments, particularly outside of the selective serotonin reuptake inhibitor (SSRI) class, which are the most commonly prescribed antidepressants. Limited knowledge beyond SSRI's can lead to challenges when caring for patients who are not seeing improvements from these treatments. Clinicians also frequently brought up the difficulty they experienced in connecting patients with psychopharmacology when their treatment needs exceeded their care providers' comfort level. Participants discussed the need for guidance when prescribing second- or third-line treatments, as there are fewer clinical standards. Further, clinicians said that they frequently worked with patients who stopped taking their prescribed medications for a variety of reasons, including stigma, negative side effects, or a lack of drug effectiveness. Based on these challenges, we talked with participants about how they currently make treatment decisions and their vision of data-driven support tools. A key result from these discussions was that effective decision support tools will need to engage with the broader healthcare system, not just an individual healthcare provider. In the remainder of this section, we describe clinicians' expectations for how DSTs may better engage with the healthcare system in order to support complex treatment decisions.

\begin{table}
  \caption{Participant details, including years of experience prescribing antidepressants, and in which of the two user studies they participated}
  \label{tab:participants}
  \begin{tabular}{p{1cm}p{1.4cm}p{1cm}p{1cm}} 
    \toprule
    ID&Experience (Years)&Study 1 (n=10)&Study 2 (n=8)\\
        \midrule
    P1&20&X&\\
    P2&19&X&X\\
    P3&<1&X&X\\
    P4&6&X&\\
    P5&5&X&\\
    P6&1.5&X&\\
    P7&9&X&X\\
    P8&16&X&\\
    P9&41&X&X\\
    P10&7&X&\\
    P11&2&&X\\
    P12&<1&&X\\
    P13&1&&X\\
    P14&1&&X\\
  \bottomrule
\end{tabular}
\end{table}


\subsubsection{Include patient preferences}
While the primary goal of this project was to address clinicians' treatment decision challenges, participants saw a clear opportunity to \textbf{use DSTs to engage patients in the decision-making process.} All of the clinicians in this study emphasized that MDD treatment decisions are a collaborative process. Though clinicians wanted to engage patients in treatment conversations, several noted the lack of available information designed for patients. Current informational resources about medications are considered to be either too simple or too complex for patients. Clinicians saw the DST output, particularly regarding personalized treatment recommendations and their associated side effects, as information that could be presented directly to patients in order to involve them in decision-making conversations, thus supporting patient-provider collaboration:

\begin{quote}
    \textit{ “If I was trying to decide between two meds and I'm talking to a patient... it's even something that you could potentially show a patient or say, "These are two choices. I think they're both maybe equally effective. This one may have more of the side effect or something."} – P8   
\end{quote}
 
Engaging patients also means providing ways to integrate patient treatment preferences into the interactive system design. Clinicians consistently said that they considered patient preferences. However, such information is not available within an electronic medical record, and therefore not accounted for within current ML models. Clinicians requested a tool for sharing and collaborating with patients, emphasizing the need for an interactive interface:

\begin{quote}
\textit{“Having an option like, patient's also worried about this and that. You can click on the two major side effects and then based on that, a specific drug will come up.”} - P3
\end{quote}

Overall, clinicians responded positively to the treatment recommendation feature. They felt such tools could remind them of treatment side-effects, and promote discussions of drug effects with patients. The emphasis on patient preferences in the treatment selection process motivated the need for interactive tools in which clinicians and patients could change the input variables. 

\subsubsection{Recommend appropriate clinical processes}
Participants frequently commented that DST predictions should be paired with recommendations for appropriate next steps in the clinical workflow, often involving other healthcare providers. Notably, while some participants helped us to connect the ML predictions to existing health system procedures, others found this task difficult. Such difficulties indicate that only displaying the prediction will likely be insufficient:


\begin{quote}
    \textit{“I think [patient dropout] is definitely an issue and it is something that I think about. I'm wondering what I would do differently if the score was higher versus lower, and if that would affect decisions or not. I'm also trying to think of what resources we would use or not in different situations.”} - P1
\end{quote}

In this case, while the participant indicated general interest in the DST recommendations, the conversation did not result in any specific conclusions on how to use the DST output to identify possible next steps. In contrast, some clinicians were able to connect the model output with appropriate and existing healthcare processes. This came up both when discussing patient dropout predictions as well as  treatment recommendations. In the context of patient predictions, clinicians shared that DSTs could recommend existing processes for patients at risk of stopping their medication:


\begin{quote}
\textit{“I feel like what that would tell me is, if there is a lower stability and higher dropout that it would be important to then involve more of a care team, rather than just say, why don't you see me in six weeks for a follow up? I would say, let me have so-and-so in my clinic call you in two weeks.”} - P7
\end{quote}

\begin{quote}
\textit{ “Especially if patients have trouble coming in, it could be longer or oftentimes that second visit is canceled. We do have some practice options for follow up. One of our pharmacists will sometimes do phone follow up and titration of these medications, so you can involve other people.”} - P2
\end{quote}

In the case of dropout prediction, participants discussed three ways in which PCPs could respond to a high dropout risk prediction: including behavioral therapy with patient counseling in the care plan, lowering the drug titration, and reducing follow-up times. Participants said that these steps could be useful for patients at risk of dropout and use resources and procedures already established within the clinic. 

We also found that while some predictions helped clinicians to identify appropriate actions and next steps, other model predictions were viewed less favorably.  While clinicians saw value in dropout risk scores, we did not receive such positive feedback regarding stability scores. Unlike the dropout scores, some clinicians did not believe that patient-level stability scores guided them towards any particular interventions or next steps, and some clinicians saw stability scores as potentially harmful for seemingly “stable” patients:

\begin{quote}
    \textit{“I don't know if it would change the initial management because I would see a patient back four to six weeks regardless, but maybe any sequential follow-up appointments I'd feel more comfortable spacing those out a little bit more.”} - P6
\end{quote}

\begin{quote}
    \textit{“I was just kind of thinking through this, how it might change my counseling when I'm prescribing the medication. I'd have to think a little bit more about that because it's not like if somebody is thought to be stable I'd want them to be suffering at home with side effects and not be reaching out to me.”}  - P4
\end{quote}

As in these examples, clinicians raised important concerns about DST recommendations. Clinicians were interested in knowing when they should follow up sooner for patients who may be at risk of discontinuing treatment, but pointed out that stability scores risk inappropriately indicating that current follow up times could be lengthened, leading to potentially negative consequences for the patient.

Clinicians also discussed expectations for direction and next steps when considering personalized treatment recommendations. Clinicians wanted tools that showed all appropriate treatment options, rather than only showing one at a time:
 
\begin{quote}
    \textit{“Rather than make me feel like there's no good option it might point me to consider something else. It may or may not be appropriate, but certainly, I would think about it I guess.” }- P2
\end{quote} 
 
As shown in the above examples, we found that clinicians expected DSTs to integrate with their existing processes and use the \textbf{DST predictions to show a path forward}. By including PCPs in design discussions, we were able to uncover which clinical processes may be most appropriate for various patients and treatment predictions. Clinicians were critical in identifying the connections between DST output and existing healthcare processes. However these connections were not apparent to all healthcare providers, indicating the decision support tools should explicitly suggest appropriate next steps within the technology design. 

\subsubsection{Understand healthcare system resource constraints}
In the interviews, we were surprised that discussions of trust in the technology were rarely initiated by participants. Probing questions revealed that PCP's limited time with patients would make in-the-moment determinations of trust nearly impossible. All participants stressed that they have limited time with each patient and must focus on understanding all of the factors that will influence their treatment decision, including the patients' medical history, symptoms, and treatment preferences. 

While participants were interested in using technology to help in their decision-making, they would not have the time in these encounters to determine if they thought a tool was trust-worthy. However, they also noted that new technology introduced into the clinic would require substantial validation through randomized controlled trials, supporting their trust in new tools. Participants said they would use it if other clinicians used it, and would make a one-time decision about if the tool was helpful:
 
\begin{quote}
    \textit{“I think the biggest thing is just getting behind how you validated your data, how you validated your model ... I don't know if you necessarily need to get into super nitty-gritty details” } - P6
\end{quote}

\begin{quote}
    \textit{“If a major medical society is sort of putting this forth, my colleagues are using it, and I hear people saying that it's that it works, then I am comfortable with it.”} - P7
\end{quote}
 
When looking at the feature importance charts, participants overall found the information unnecessary in determining how they will care for a patient, and ill-fitted for their short patient appointments:


\begin{quote}
   \textit{“[The features] just feel a little random, these things. Again, I don't know if it would help me. I'm just not sure how it's going to help me change what I'm going to do.”} - P8
\end{quote}

Thus, we found that conversations of trust were not typically initiated by participants because \textbf{participants expected that trust in the technology will not be decided at each decision point.} This result contrasts with  work in the field of explainable AI, which often looks at designing explanations for each prediction or recommendation. Based on this feedback, we see that data validation procedures should be findable, but not forced upon providers who are already focused on understanding many complicated facets of a patients' history, health status, and treatment needs. 

\section{Prototype Redesign}
We redesigned the DST prototype to operationalize the guidelines established in study 1. Table \ref{tab:design} describes the features we included in the prototype. We first considered ways in which we could better account for patient preferences, as clinicians all described the treatment selection process as a mutual and collaborative decision. Therefore, we made the treatment recommendations more interactive, allowing clinicians to edit which aspects of a patient's medical history are being used to drive the recommendations. For example, a clinician could select `fatigue' if the patient was experiencing this symptom, and antidepressant treatments that are recommended for treating fatigue (and do not have a negative interaction with other patient symptoms) would appear as favorable. Educating patients about the potential negative side effects of the treatments was also an important aspect of clinicians' work, as this communication could help in collaborative decision-making and help reduce the risk of the patient discontinuing the treatment. We therefore  display on hover all of the rules associated with a treatment. By displaying all of the rules for a drug, both the PCP and patient can better understand the potential effects and ensure that the treatment they select is appropriate for the patient. Given the important role patients play in the treatment selection process, we also acknowledge the importance of creating patient-facing tools to support education and decision-making. For this study, we focused specifically on clinician-facing tools, but plan to look into designing multi-user systems for patient-provider collaboration in the future.

\begin{table*}
  \caption{Summary of changes made to the design of the MDD decision support prototype based on study 1 findings}
  \label{tab:design}
  \begin{tabular}{p{4.5cm}p{9cm}} 
    \toprule
    \textbf{\textit{Include patient preferences}} & \textbf{Support patient-provider communication, address missing variables}
    \begin{itemize}
        \item Make treatment recommendations interactive, so that clinicians and patients may edit the input variables based on changes to a patient's medical history or side effect preferences
        \item When hovering over a treatment, show all potential side effects for that antidepressant, in order to foster communication and education of potential medication effects
    \end{itemize}\\
    \midrule
    \textbf{\textit{Recommend appropriate \newline clinical processes}}& 
    \textbf{Show a path forward, provide actionable information}
    \begin{itemize}
        \item For patients with a dropout risk prediction in the top quartile, present recommended next steps based on clinicians' suggestions
        \item Allow for viewing and comparing multiple antidepressant options
    \end{itemize}\\
    \midrule
    \textbf{\textit{Understand system constraints}}&
    \textbf{Do not require determination of trust at every decision point}
    \begin{itemize}
        \item Refocus from model features to model validation process
        \item Present an overview of all model validation steps in a single screen that is accessible from the main interface, but not combined with patient details
    \end{itemize}\\
  \bottomrule
\end{tabular}
\end{table*}

Our second goal was to connect the DST predictions with appropriate clinical processes. Based on the discussions from the first study, in which participants expressed interest in patient dropout scores, we aimed to make this more prominent in the second iteration. We found through discussions with our clinical collaborators that the dropout prediction was difficult to understand without knowing the distribution of predictions across all patients. We included a graph to help visualize this distribution and support further conversation of information needs. We also included next steps PCPs could consider for high dropout risk predictions. As mentioned previously, these steps included lowering the medication titration, scheduling earlier follow up visits with the patient, and making sure patients were set up with additional behavioral therapy.  

We also altered the interface for the treatment recommendations. Rather than asking clinicians to search for a specific treatment, we used a matrix design that allows clinicians to compare antidepressants side by side. Participants indicated that such comparisons were important for helping them consider all of the possible treatment options. We selected the matrix design in order to display a large amount of information in a glanceable display. Also, PCPs currently use a publicly available table that includes information about antidepressant treatments. The matrix design mirrored the table format that PCPs currently use, while adding the needed interactivity.

Finally, we aimed to design the prototype to work within the time-constraints of the medical system. We first replaced the feature-importance explanations with information about the model validation process. To view this information, we added a link labeled \textit{how dropout is calculated} below the dropout prediction. The link leads to a page that lists the steps used to validate the model. We expect that this page will expand with each validation study, including clinical trial protocols, results, and publications. Clinicians did express interest in how the model was validated and the results of any randomized trials or other validation studies of the technology. By making these details easily findable, our intention was to make this information accessible, while also respecting PCPs' limited time with patients. We aimed to make validation data available, without distracting from the information most critical to the decision-making process.

\begin{figure*}[]
  \centering
  \includegraphics[width=.95\linewidth]{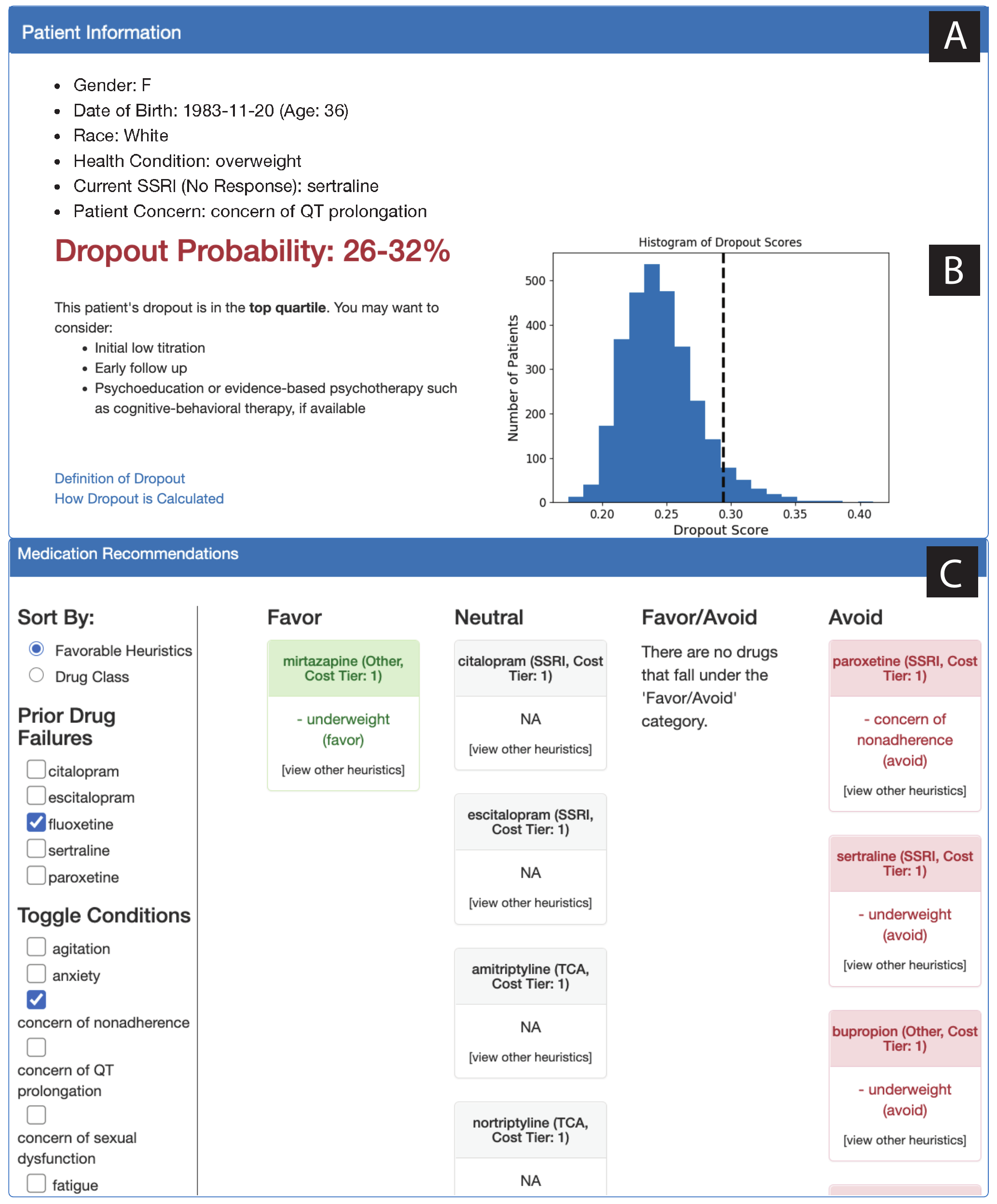}
  \caption{Features included in the prototype redesign (from top to bottom): (A) patient information, (B) dropout score with links to further information about how dropout is defined and validation studies conducted on the tool, (C) interactive personalized treatment recommendations.}
  \Description{The figure shows three features included in the prototype. The top image includes a list of patient information. The middle image states a droput probability range of 26-32\% with a bar chart showing that this range is in the top-quartile of patient dropout risk predictions. The bottom feature is an interactive treatment selection tool in which providers may toggle patient conditions and antidepressant treatments will organize in four columns: favor, neutral, favor/avoid, and avoid.}
  \label{fig:two}
\end{figure*}

\section{USER STUDY 2: PROTOTYPE FEEDBACK}
\subsection{Methods}
\subsubsection{Participants and recruitment}
For this study we again worked with PCPs who currently prescribe antidepressant treatments. To recruit participants we worked with the same academic medical center and used the same recruitment method as with the first user study. We invited both new participants as well as those who participated in the prior study. We provided each participant with a \$20 Amazon gift card to thank them for their time.

\subsubsection{Study Design}
This study was approved by the Harvard Institutional Review Board. Each study session lasted 30 minutes and was conducted remotely using Zoom. During the study session, participants were provided links to the prototype, asked to share their screen, and were able to interact with the prototype freely. After a participant was presented with the link, we asked the following questions in order to guide discussion and feedback:
\begin{enumerate}
    \item Imagine this patient is sitting in front of you, how would you make a treatment decision?
    \item What helped you make a decision?
    \item Did anything detract from making a decision? Or your confidence in the decision?
    \item If you were going to design this tool for a colleague, what would they need to make a decision? What would you change?
\end{enumerate}
Due to our focus on walking through a decision process that replicates real-world decision-making we worked with experts in clinical psychology and psychopharmacology to create a realistic patient scenario. We presented clinicians with a short summary of essential patient information, including age, gender, relevant comorbidities, and a prior ineffective SSRI trial, as shown in figure \ref{fig:two}A.

\subsubsection{Data Analysis}
Similar to study 1, all sessions were audio recorded and transcribed. To analyze the data, we continued to use a grounded theory approach, \cite{Charmaz2012}, first using an iterative inductive analysis to establish themes within the data. As in study 1, the first author open coded the transcripts and identified an initial set of themes. The research team then met to discuss themes and associated data segments, discuss discrepancies, and amend theme definitions. The first author then re-coded the data using the established themes and met with the research team to validate the final set of themes. During this process, we found many similar themes to study 1. Therefore we also ran a deductive coding process \cite{Elo2008}, using the codes from the first user study in a subsequent analysis to identify thematic overlap, allowing us to assess specific feedback related to the study 1 guidelines.

\subsection{Findings}
Eight PCPs enrolled in this user study (six physicians, two residents, and one nurse practitioner), four of whom also participated in the first study. Participants in this study had an average of 9.4 years of experience prescribing antidepressants (SD=14.3). While we did not directly ask about the results of study 1, we found that much of the feedback re-emphasized the themes we previously discussed, helping to validate our previous findings and our approach to operationalizing those guidelines. The ability to interact with a high fidelity prototype also led participants to discuss new opportunities and challenges.

\subsubsection{Feedback on Study 1 Guidelines}
Much of the prototype feedback re-emphasized the lessons we learned in the first user study. Participants discussed ways in which the prototype successfully met their needs and opportunities for the system to further align with their expectations.

Participants positively responded to the ways in which we integrated clinical processes by displaying relevant actions. Participants commented that the recommended steps associated with dropout risk predictions were actionable and aligned with what they typically do for patients when they are concerned about their response to treatment:

 
\begin{quote}
     \textit{“One of the things that is the hardest about depression and anxiety is that patients who have really bad symptoms tend to not follow up or tend to not follow through with therapies just due to the nature of their disease. And so going into a room and already knowing, is there a high chance that this patient may fail on this therapy or may not adhere to this therapy, and going into the room knowing that this is someone who I need to talk to a little bit more or I need to follow up with a little bit more or who I need to schedule really close appointments for. I think that's probably the biggest help that you can offer.”} - P11
 \end{quote}

In addition to the positive feedback, some participants mentioned opportunities to continue to integrate the DST with existing clinical processes. For example, multiple clinicians discussed the benefit of connecting treatment recommendations with their prescription systems:
 
\begin{quote}
\textit{“I guess if I were using this in practice, I would probably click to see if prescribing information or something like that came up as a next step.”} - P2
\end{quote}
 
Clinicians also responded positively to the ability to edit patient conditions to see how the inclusion or exclusion of various conditions changed treatment recommendations, allowing them to incorporate patient concerns: 
 
\begin{quote}
 \textit{“You can sort of click this and it helps to think about if I add poor concentration, if maybe that's another side effect she's having and it kind of modifies the medication regimen based off of that and bupropion is still appropriate for that. So I think that toggling conditions is super helpful."} - P11
\end{quote}
 
Clinicians also suggested additional conditions that should be included, such as pregnancy and suicidality, as these will influence their treatment decision. Clinicians also discussed ways in which the prototype could further support patient communication. Participants did respond that the treatment recommendations could facilitate patient counseling, as expressed in study 1. Even further, some clinicians saw an opportunity to use the information from the DST to create educational sources for patients:
 
\begin{quote}
\textit{“It would be nice if, after you make a selection, like say I click, okay, we're going to go with bupropion, if there was a patient-friendly educational handout that would just say, we're starting you on bupropion. As a reminder side effects to expect, side effects that are less common. Just something that I could give the patient so that they remember why we went on this and what might be normal, because a lot of the time I'll end up writing that down. But if it's already in here and I can hit print, that might be useful."} - P14    
\end{quote}

Finally, a goal with this prototype was to allow clinicians to establish trust within the tool while being more mindful of their time constraints. We found that, similar to study 1, discussions of trust were infrequently initiated by participants, but the link to model validation did pique participants' interest. Interacting with this part of the prototype led to conversations about the types of validations that clinicians would expect to see:
 
\begin{quote}
\textit{“If you could show that patients have a better response to treatment by use of the algorithm, that would be amazing. If you can show that patients actually are more likely to adhere to treatment, that would be important as well, or that patients are less likely to develop adverse side effects that leads to stopping medications. It would be nice to do a trial with outcomes like that.”} - P9    
\end{quote}

These discussions reemphasized the type of validation methods that would help clinicians' to establish trust. Overall, we found that participants positively responded to the design concepts, reemphasized the expectations established in the previous study, and discussed ways in which the system could further align with these expectations in subsequent designs. As we discuss in the next section, interacting with the prototype also revealed a new way in which DSTs must engage with the broader healthcare system.

\subsubsection{Engaging with Domain Knowledge}
While participants responded positively to the overall prototype design, we found that when the recommendations diverged from clinical knowledge or guidelines participants became confused and would often abandon the recommendation. There were two ways in which the the prototype differed from clinicians' expectations. First, the system predicted a high dropout risk for a patient who participants would not typically assume to be high-risk patients:

\begin{quote}
 \textit{“To me, when I think of someone who's a high risk of dropout, it's like a person with substance abuse or with bipolar disorder. Like, okay, obviously, they're not going to show up. There's a high chance they won't adhere to it. But for this person who seems like a relatively bread and butter, middle-aged, healthy person, the fact that she's on the higher end of dropout is kind of eye opening to me, and it almost makes me wonder then, who is on the lower end of dropout?”} – P7   
\end{quote}

Surprising predictions, such as this one, led to greater interest in the depression score and how it was calculated. For example, as in the quote below, participants' next step was to understand what factors were leading to this unexpectedly high dropout prediction:

\begin{quote}
\textit{“And I have to like, look more into this, but the dropout probability, is it because of the side effects that you're dropping out? Or is it because the medication is not effective.”} - P3
\end{quote} 
 
Here, we see that when faced with unexpected model output that contrasted with participants' expectations or mental model, participants found it challenging to identify appropriate next steps. We also found that treatment recommendations at times contrasted with participants' expectations. Specifically, treatments that were listed as either favorable or neutral included medications that clinicians would not typically prescribe to patients. In some cases, clinicians indicated that seeing a surprising treatment recommendation would give them pause and lead to more reflection about the patient's case and medication needs:

\begin{quote}
\textit{“So truthfully, I would take a step back because it's not that common that nortriptyline is a medication I think about as a first or even a second or third line agent, unless they have other conditions that I know [tricyclic antidepressants] can treat. So I would really take a step back and think about the patient's pain. Do they have really bad migraines, that I think will get significant benefit from the TCAs. It would definitely give me pause if that was the most favorable medication to come up as a suggestion on this.”} – P11
\end{quote}

The examples we presented in this section describe reactions to scenarios in which the output of the machine learning model contrasts from clinical experiences or standards of care. Importantly, as new machine learning models continue to advance, so may the opportunities for model output to contrast with clincians' expectations or existing care standards. Our findings reveal a need for researchers to consider how to \textbf{adapt DST system designs for instances in which the machine learning model output constrasts with existing domain knowledge.}

\section{DISCUSSION}
Through conversations on their expectations for AI support, clinicians revealed a number of critical aspects of the sociotechnical healthcare system that need to be considered in the design of novel decision support tools. Specifically, our work highlights the importance of including patient preferences, recommending clinical processes, understanding system constraints, and engaging with domain knowledge. In the remainder of this section, we discuss how the prototype feedback reveals lessons for how we design AI for healthcare systems and we reflect on the implications of this work for HCI research.

These empirical implications provide concrete implications to the specific context of MDD. We believe the lessons learned from this work may be transferable to healthcare systems and process that share characteristics that were emphasized by the participants in this work: 1) Patients are deeply involved in treatment decisions, 2) Providers have short and infrequent appointments with patients. In particular, we expect these results will be useful for designing ML-embedded DSTs for other primary care decisions. Future work should look at how needs and expectations change for medical specialists, healthcare systems in which patients are unwilling or unable to participate in care decisions, and for patients directly.

\subsection{Creating multi-user systems for collaborative decision-making}
The clinicians who participated in this study encouraged us to use DSTs to foster greater collaboration with patients. While AI research continues to focus on improving information accuracy for clinicians, modern clinical care aspires toward shared decision-making with patients and clinicians working together to make decisions~\cite{Charles1997,Makoul2006}. Consequently, \textbf{clinicians' feedback challenged the idea that AI-driven innovations in healthcare will be single user systems.} We expect that designing multi-user systems that engage patients and their healthcare providers may have several benefits. First, such tools can help patients have a greater voice in their healthcare decisions. Health tools that directly interact with patients can help promote patient activation by increasing access to important health information and providing new ways to engage in their health care \cite{Solomon2012, Schnock2019}. Studies have connected increased patient activation to improved healthcare experiences and health outcomes~\cite{Hibbard2013}. A recent study also found that including patient-facing DSTs can improve clinicians' adherence to recommended practices when compared to DSTs that were only clinician-facing~\cite{VandeVelde2018}. Finally, we believe multi-user systems can support time management during clinical encounters. DST's will likely be able to communicate healthcare options quickly to patients, and may provide tailored educational materials that are continuously available to patients. Such technological support may address both the time constraints in primary care settings and the cognitive constraints of the patient, who can experience information overload in clinical encounters~\cite{Khaleel2020}. Yet, very few studies have looked at creating AI tools for patients~\cite{Karhade2019}. In an initial attempt to promote more inclusion of patients' preferences, we used interaction to help tailor treatment recommendations. However, we see this as a small step towards a larger problem. In the future, we intend to co-design such systems with patients directly, in order to amplify their voices in their own care decisions. 

\subsection{Connecting DSTs to Existing Healthcare System Processes}
We see a clear need for DSTs to explicitly draw the connections between the model output and actionable next steps. A consistent theme within this study was that clinicians wanted tools that provided actionable interventions, connecting predictions to appropriate clinical processes. Often, these processes involve additional healthcare providers. Thus, decisions in healthcare, such as treatment selection, are not siloed tasks. Rather, these healthcare decisions affect many other aspects of care. For MDD, this meant connecting patients with behavior therapy, pharmacology, and nurses who could follow up with them and track their progress. \textbf{The ability for some clinicians' to connect DST output to existing clinical processes  demonstrates the benefit of integrating participatory design methods into DST development workflows.} 

We also found that clinicians were thoughtful in considering the possible adverse effects of DST predictions. This became clear as clinicians considered the potential effects of stability and dropout predictions. While clinicians saw value in tailoring care for patients with high dropout risk, some clinicians were wary of stability scores. Some clinicians commented that they were unable to identify a clear next step for patients with high stability scores, while others were concerned that these patients would not receive the needed attention. We have seen several examples in recent years of AI predictions leading to biased or unfair behaviors \cite{Green2019a, Obermeyer2019}. Engaging clinicians or other target users of AI tools in design fiction methods~\cite{Noortman2019} may be another useful step in the AI design process.

\subsection{Designing for Resource Constraints}
While explaining black box models is a consistent theme in AI work, we need best practices for adapting the design of DSTs for time-constrained environments. In the case of MDD, and primary care settings more generally, time constraints will consistently need to be considered in the design of any novel health tools. Time constraints have been cited as the most common barrier to the adoption to new decision support tools \cite{Devaraj2014}.

Designing for fast-paced, time-constrained work environments has important design implications, particularly in the context of current explainable AI research. In our work, we found that due to time limitations, clinicians wanted to determine their trust in the technology one time, rather than at each decision point. Therefore, \textbf{clinicians wanted DSTs to display the evidence-based methods used to validate the tool (such as randomized controlled trial results), rather than individual explanations that focus on model features.} Prior work has also noted that clinicians wanted ML tools to more closely align with evidence-based medicine methods \cite{Tonekaboni2019}. In our iterative design, our shift from model features as explanations to tool validation steps helped to better reflect the evidence-based medicine process.

Recent literature in explainable AI has made important progress towards improving transparency of AI models by creating interpretable explanations for the model's output \cite{Wang2019}. These explanations can influence decision-makers understanding of the model and perceived fairness of these tools \cite{Dodge2019}. However, in high stakes or time critical environments, this process places time and mental burden on users. In the case of primary care our results indicate that explanations for each model prediction would be unusable due to the limited time clinicians have with each patient. Our results therefore align with recent discussions of the potential problems of using explainable AI in clinical settings \cite{Rudin2019}. However, without the time to review an AI prediction or recommendation in detail, greater responsibility needs to be placed up front to determine when the system is likely to err, and when the tool should perhaps not be shown altogether.

\subsection{Adapting decision support for contrasting information}
We found that when the DST output did not align with clinical knowledge or guidelines, clinicians were left confused, with most opting to abandon the recommendation. While explanations to support transparency at each decision point were seen as unhelpful in general, more details in the case of surprising predictions were viewed more favorably. In these cases, clinicians requested information about causal factors that would allow them to intervene appropriately. This is notably distinct from the feature-importance explanations we produced in study 1, which included correlated features produced by the model. Prior studies have also found that feature-based explanations were inadequate for helping clinicians identify appropriate interventions \cite{Yang2019a}.

We do not yet have established best practices for dealing with contrasting information, but helping clinicians identify the best way to proceed is critical. Contrasting information in medicine can result in clinical uncertainty and adverse effects for patients \cite{Carpenter2016}. The introduction of AI in clinical settings will likely increase the prevalence of contrasting information, as ML models trained on vast data sets have the promise to uncover nuanced relationships that are not encoded in 
existing medical training, guidelines, or clinician's expectations. We see this as an important opportunity for future work to develop best practices for cases in which DST recommendations diverge from domain knowledge, and show why divergence is happening in a way that is both understandable to the user and presents actionable next steps.

Based on our results, we see an opportunity to \textbf{present on-demand explanations as differentials from existing clinical guidelines.} This is consistent with the emerging understanding in the explainable AI community that in human-human discourse explanations are typically \em contrastive\em~\cite{miller2019explanation,Pradier2021}. Rather than providing all evidence in support of a recommendation, such explanations could visualize how and why a machine learning model output is diverging from existing guidelines or expert knowledge. This will require machine learning models to robustly reason with existing mental models of the users. In healthcare, the use of standardized clinical guidelines (e.g.,~\cite{macqueen2017systematic,Kennedy2016} for MDD treatment selection) means that aspects of these mental models are established, encodable, and may therefore be used in both the development of the machine learning models and the design of DST interfaces. Ehsan and Riedl have also suggested that allowing users to voice skepticism in AI models may afford new interactions that encourage users to consider the limitations of the technology~\cite{Ehsan2020}. In the context of healthcare decisions, allowing clinicians and patients to voice skepticism and highlight surprising DST outputs may allow the underlying models to adapt as medical guidelines continue to evolve. A critical area for future work is designing tools which allow users to identify contrasting information in time-constrained environments and determine how to best proceed.

\section{LIMITATIONS}
The application of machine learning in healthcare brings numerous practical, ethical, and legal issues. Here we address one challenge in implementing these tools in the real world---considering the broader healthcare system. In this study we used an iterative design process to guide discussions of clinician expectations. We do not expect that this is the optimal visualization and much work is needed in the area of data visualization to represent both the distributions of ML predictions, as well as the uncertainty within the model. Further, in this study, we focused solely on the perspective of clinicians. While we see this as an important first step, next-step studies should engage with other stakeholders, such as patients, nurses, pharmacists, and therapists.

\section{CONCLUSION}
In this paper, we report on aspects of the healthcare sociotechnical system that should be considered in the design of machine learning decision support tools. Based on co-design studies with primary care providers, we identified four important aspects of the care system that influence how we design decision support tools for real-world use. These factors include patients' preferences, clinical processes that often include multiple healthcare providers, the constraints of the healthcare system, and existing domain knowledge. We posit that by making these aspects of healthcare central to the design of DSTs, we may develop tools that are capable of supporting the collaborative nature of healthcare, identifying potential adverse events caused by ML predictions, working within time-critical environments, and recognizing conflicting information. We do not expect that the sociotechnical factors discussed here represent the full set of sociotechnical considerations that need to be included in AI design. Rather, we present this as an initial step in a broader research agenda determining how the new wave of intelligent systems must account for the complexity of medical work.






\begin{acks}
This research was supported in part by the Harvard Data Science Initiative. We thank all of the participants for taking time to share their insights and expertise.
\end{acks}

\bibliographystyle{ACM-Reference-Format}
\bibliography{output}










\end{document}